\documentstyle[preprint,aps,eqsecnum,epsf,floats]{revtex}
\begin{document}
\tighten

\def\bfl{{\bbox \ell}}
\def\bull{\vrule height .9ex width .8ex depth -.1ex}
\def\MeV{{\rm MeV}}
\def\GeV{{\rm GeV}}
\def\Tr{{\rm Tr\,}}
\def\nrcpt{NR\raise.4ex\hbox{$\chi$}PT\ }
\def\ket#1{\vert#1\rangle}
\def\bra#1{\langle#1\vert}
\def\ltap{\ \raise.3ex\hbox{$<$\kern-.75em\lower1ex\hbox{$\sim$}}\ }
\def\gtap{\ \raise.3ex\hbox{$>$\kern-.75em\lower1ex\hbox{$\sim$}}\ }
\def\abs#1{\left| #1\right|}
\def\CA{{\cal A}}
\def\CC{{\cal C}}
\def\CD{{\cal D}}
\def\CE{{\cal E}}
\def\CL{{\cal L}}
\def\CO{{\cal O}}
\def\CZ{{\cal Z}}
\def\bvert{\Bigl\vert\Bigr.}
\def\pds{{\it PDS}\ }
\def\ms{MS}
\def\ddq{{{\rm d}^dq \over (2\pi)^d}\,}
\def\ddqm{{{\rm d}^{d-1}{\bf q} \over (2\pi)^{d-1}}\,}
\def\bfq{{\bf q}}
\def\bfk{{\bf k}}
\def\bfp{{\bf p}}
\def\bfpp{{\bf p '}}
\def\bfr{{\bf r}}
\def\dtr{{\rm d}^3\bfr\,}
\def\bfx{{\bf x}}
\def\dtx{{\rm d}^3\bfx\,}
\def\dfx{{\rm d}^4 x\,}
\def\bfy{{\bf y}}
\def\dty{{\rm d}^3\bfy\,}
\def\dfy{{\rm d}^4 y\,}
\def\dfq{{{\rm d}^4 q\over (2\pi)^4}\,}
\def\dfk{{{\rm d}^4 k\over (2\pi)^4}\,}
\def\dfl{{{\rm d}^4 \ell\over (2\pi)^4}\,}
\def\dtq{{{\rm d}^3 {\bf q}\over (2\pi)^3}\,}
\def\dtk{{{\rm d}^3 {\bf k}\over (2\pi)^3}\,}
\def\dtl{{{\rm d}^3 {\bfl}\over (2\pi)^3}\,}
\def\dt{{\rm d}t\,}
\def\frac#1#2{{\textstyle{#1\over#2}}}
\def\darr#1{\raise1.5ex\hbox{$\leftrightarrow$}\mkern-16.5mu #1}
\def\){\right)}
\def\({\left( }
\def\]{\right] }
\def\[{\left[ }
\def\si{{}^1\kern-.14em S_0}
\def\siii{{}^3\kern-.14em S_1}
\def\diii{{}^3\kern-.14em D_1}
\def\p0iii{{}^3\kern-.14em P_0}
\def\fm{{\rm\ fm}}
\def\MeV{{\rm\ MeV}}
\def\CA{{\cal A}}
\def\Czzm{ {\cal A}_{-1[00]} }
\def\Cttm{{\cal A}_{-1[22]} }
\def\Ctzm{{\cal A}_{-1[20]} }
\def\Cztm{ {\cal A}_{-1[02]} }
\def\Czzz{{\cal A}_{0[00]} }
\def\Cttz{ {\cal A}_{0[22]} }
\def\Ctzz{{\cal A}_{0[20]} }
\def\Cztz{{\cal A}_{0[02]} }


\def\spzz{ {Y_{sp0}^{(0)} }}
\def\spzzB{ {Z_{sp0}^{(0)} }}
\def\spzo{ {Y_{sp0}^{(1)} }}
\def\spzt{ {Y_{sp0}^{(2)} }}
\def\ppspz{ {y^{sp0}_0 }}
\def\ppspt{ {y^{sp0}_2 }}
\def\ppspB{ {z^{sp0}_2 }}

\def\Ames{ A }  

\newcommand{\eqn}[1]{\label{eq:#1}}
\newcommand{\refeq}[1]{(\ref{eq:#1})}
\newcommand{\eq}{eq.~\refeq}
\newcommand{\eqs}[2]{eqs.~(\ref{eq:#1}-\ref{eq:#2})}
\newcommand{\eqsii}[2]{eqs.~(\ref{eq:#1}, \ref{eq:#2})}
\newcommand{\Eq}{Eq.~\refeq}
\newcommand{\Eqs}{Eqs.~\refeq}


\def\logone{{\log\left[ {(m_\pi + 2 \gamma)^2 \over 
        (m_\pi+2 \gamma)^2 + {|{\bf k}|^2 \over 4}}\right]  }}
\def\atanfour{{\tan^{-1}\left( {|{\bf k}| \over 4 \gamma}\right)  }}
\def\atantwo{{\tan^{-1} \left({|{\bf k}| \over 2 \gamma}\right) }}
\def\atanthree{{\tan^{-1} \left({|{\bf k}| \over 2
        (m_\pi+ 2 \gamma)}\right) }}

\def\Journal#1#2#3#4{{#1} {\bf #2}, #3 (#4)}

\def\NCA{\em Nuovo Cimento}
\def\NIM{\em Nucl. Instrum. Methods}
\def\NIMA{{\em Nucl. Instrum. Methods} A}
\def\NPB{{\em Nucl. Phys.} B}
\def\NPA{{\em Nucl. Phys.} A}
\def\PLB{{\em Phys. Lett.}  B}
\def\PRL{\em Phys. Rev. Lett.}
\def\PRD{{\em Phys. Rev.} D}
\def\PRC{{\em Phys. Rev.} C}
\def\PRA{{\em Phys. Rev.} A}
\def\ZPC{{\em Z. Phys.} C}
\def\PREP{{\em Phys. Rep.}  }
\def\ANN{{\em Ann. Phys.} }
\def\SCI{{\em Science} }

\preprint{\vbox{
\hbox{ NT@UW-99-36}
\hbox{ DUKE-TH-99-191}
}}
\bigskip
\bigskip

\title{The Anapole Form Factor of the Deuteron}
\author{Martin J. Savage} 
\address{Department of Physics, University of Washington, Seattle, WA 98195.
  \\ {savage@phys.washington.edu} \\
and \\
 Jefferson Lab., 12000 Jefferson Avenue, Newport News, \\
Virginia 23606.}
\author{Roxanne P. Springer\footnote{\rm  
On leave from the Department of Physics,   Duke University, Durham NC 27708.
    }}
\address{Institute for Nuclear Theory, University of Washington, Seattle, WA
98195 \\ {rps@phy.duke.edu }}
\maketitle

\begin{abstract}
We find the leading order result for the anapole form factor of the
deuteron using effective field theory. 
The momentum dependence of the anapole form factor is different
from that of the matrix element of the strange currents in the 
deuteron, which may provide a way for disentangling these  two
competing 
effects when analyzing
parity violating electron-deuteron scattering experiments.
We give closed form expressions for both the form factor
and integrals often encountered in the NN effective field theory. 
\end{abstract}

\section{Introduction}

Several open questions in nuclear parity violation\cite{PVprobs,PVprobsb}
prompt further
investigation of parity violating observables in the simplest nuclear
system.  
Recently, a precise measurement of the parity violating angular asymmetry
in $\overrightarrow np\rightarrow d\gamma$ was proposed\cite{npprop}
in order to tightly constrain the weak 
one-pion-nucleon coupling constant, $h_{\pi NN}^{(1)}$.
In addition, 
MIT-Bates has a dedicated program to measure parity violation
in electron scattering on the deuteron\cite{bates}, which  depends
upon both the strangeness content of the deuteron, and on 
weak nonleptonic interactions which generate an
anapole interaction.   
Both measurements will constrain  $h_{\pi NN}^{(1)}$,
free from  finite density effects that contaminate extractions
from measurements in light and heavy nuclei.

For low momentum scattering, an effective field theory 
treatment\cite{Weinberg1,KoMany,Parka,KSWa,KSW,KSW2,CoKoM,DBK,cohena,Fria,Sa96,LMa,GPLa,Adhik,RBMa,Bvk,aleph,Parkb,Gegelia,steelea,CGSS,SSana,binger} 
of the deuteron is both
systematic and predictive\cite{KSW2,CGSS}.  The language and notation used
in this paper is based upon refs.\cite{KSW} and \cite{SSana}.  
Here we generalize the calculation of the deuteron anapole
moment\cite{SSana} by
retaining the dependence upon the momentum transfer
from the virtual photon, thereby
determining the full anapole form factor at leading order (LO) in the 
effective field theory expansion.

If sufficient data can be taken at $Q^2$ low enough for the
effective field theory to be valid, the momentum dependence of the
parity violation from electron scattering on the deuteron
should follow, up to higher order corrections, the form we
obtain.
In that case, it may be possible to extract not
only the strange axial matrix element of the nucleon, but also the
(presently controversial) value of the parity violating pion-nucleon
coupling.

\section{Effective Lagrangian}

We give here the effective Lagrangian needed
for this calculation\cite{KSW,SSana}.  
A discussion of the power counting used to
determine the order at which each diagram contributes is  presented there.
Both strong and weak interactions are required 
for the leading order anapole form factor.

The strong interactions of the pions and nucleons
are described by 
\begin{eqnarray}
{\cal L}_{strong} & = &  {f^2\over 8} Tr D_\mu \Sigma
D^\mu \Sigma^\dagger
+ N^\dagger \left(i D_0 + {{\bf D}^2\over 2M_N}\right) N + {ig_A\over
2} N^\dagger {\bf \sigma} \cdot (\xi {\bf D} \xi^\dagger - \xi^{\dagger} {\bf
D} \xi)N \nonumber \\
& + & {e \over 2 M_N} N^\dagger(\kappa_0 + \kappa_1 \tau_3) \sigma \cdot
{\bf B} N -
C^{(\siii)}_0
(N^T P_{i,0} N)^\dagger(N^T P_{i,0} N)
\ +\ \cdots
\ \ \ ,
\end{eqnarray}
where the pion fields are contained
in a special unitary matrix,
\begin{equation}
\Sigma = \xi^2 = \exp {2i\Pi\over f},\qquad \Pi =
\left(\begin{array}{cc}
\pi^0/\sqrt{2} & \pi^+\\ \pi^- & -\pi^0/\sqrt{2}\end{array} \right),
\end{equation}
with $f=132\ \MeV$.  $N$ is the isospin doublet of nucleon fields, $M_N$ is
the nucleon mass, $g_A~\sim~1.25$ is the axial coupling constant,
$D^\mu$ is the covariant derivative, 
and the  ellipses represent operators involving more insertions of the light
quark mass matrix, meson fields, and spatial derivatives.  The isoscalar
and isovector magnetic moments are $\kappa_0=0.44$ and $\kappa_1=
2.35$ in units of nuclear magnetons.
$P_{i,0}$ is the spin-isospin projector for the $\siii$ channel, defined by
\begin{eqnarray}
P_{i,0} & \equiv & {1\over\sqrt{ 8} } \sigma_2\sigma_i\ \tau_2\ , 
\qquad \Tr P_{i,0}^\dagger
P_{j,0} = {1\over 2} \delta_{ij}\ ,
\end{eqnarray}
The coefficient $C_0^{(\siii)}  =  -5.51\fm^2 $ has 
been determined from fits to the S-wave phase shifts in
the $\siii$ channels~\cite{KSW}, renormalized at $\mu=m_\pi$
in the power divergence subtraction scheme (PDS)\cite{KSW}.

The leading order weak interaction Lagrange density is
\begin{eqnarray}
    {\cal L}^{\Delta I=1}_{\rm weak} & = &
i\ h_{\pi NN}^{(1)}\ \pi^+\ p^\dagger n\ +\
  {\rm h.c.}\ + \cdots
    \end{eqnarray}
where the ellipses represent higher order terms  
in the chiral and momentum expansion.

The value of the parity violating pion-nucleon coupling, $h_{\pi NN}^{(1)}$,
is controversial. Theoretical treatments provide qualitative
ranges, but experimental results are in 
disagreement\cite{PVprobs,PVprobsb,Fluorine,anawei}.  
However, if we take the generally accepted size of 
$h_{\pi NN}^{(1)} \sim 10^{-7}$ \cite{DDH,adelhax,MMa,DZ,KSa,HHpv}, 
then $h_{\pi NN}^{(1)}$ is responsible
for the leading order parity violating effects in the deuteron.
Strong interaction effects in the deuteron are dominated by
$C_0^{(\siii)}$, with higher order corrections coming
from pion exchange and contact terms multiplying operators with
two derivatives or quark mass insertions.  
In the KSW power counting (outlined in \cite{KSW}),
strong interaction pions are treated
in perturbation 
theory, and are sub-leading compared to the four-nucleon contact interactions.
In the calculations of deuteron observables and interactions that have been 
computed with KSW power counting, there are no indications that the pionic
contributions are larger than naively expected.
Calculations at next-to-next-to leading order (NNLO)  
(for recent progress in this area see \cite{FMSa,MSa})
will allow more definite conclusions
about the convergence of the theory to be made\cite{CHa}.


\section{The Anapole Form Factor of the Deuteron}

The anapole form factor of the deuteron, $A_D(|{\bf k}|)$,
describes its  spin-dependent coupling to the
electromagnetic field for  arbitrary momentum transfer, 
defined through
\begin{eqnarray}
  {\cal L} & = & i A_D(|{\bf k}|)\ {1\over M_N^2}\
  \epsilon_{abc} D^{a\dagger} D^b  \partial_\mu F^{\mu c}
  \ \ \ ,
\end{eqnarray}
where  ${\bf k}$ is the three momentum of the photon, 
$F^{\mu c}$ is the electromagnetic field
strength tensor, and
$i \epsilon_{abc} D^{a\dagger} D^b$ is the deuteron spin operator.
The equations of motion for the electromagnetic field allow this interaction
to be written in terms of local four-fermion interactions, and therefore
this does not correspond to a coupling between a nucleon
and an on-shell photon.

\begin{figure}[t]
\centerline{\epsfxsize=4.0in \epsfbox{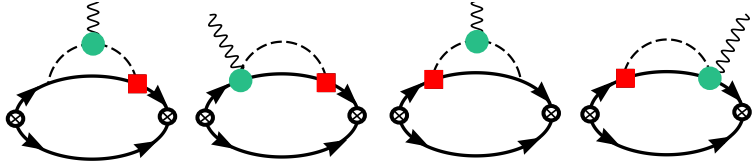}}
\noindent
\caption{\it 
Leading order diagrams contributing to the deuteron anapole  
form factor.  These are the single nucleon contributions, written 
  as $A_N(|{\bf k}|)$  in the text.    
  The crossed circles represent operators that create  
  or annihilate two nucleons with
  the quantum numbers of the deuteron.
  The solid square is the weak operator with 
  coefficient $h^{(1)}_{\pi NN}$, and the solid circle is the minimal 
  coupling
  to the electromagnetic field.
  Wavy lines are photons, solid lines are nucleons, 
  and dashed lines are mesons.
  }
\label{fig:anasn}
\vskip .2in
\end{figure}
\begin{figure}[t]
\centerline{\epsfxsize=4.0in \epsfbox{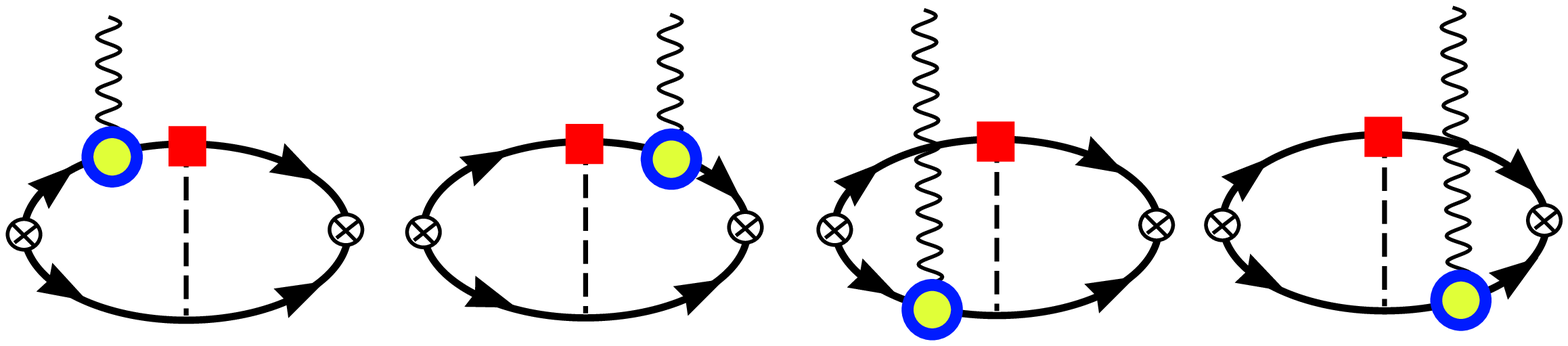}}
\noindent
\caption{\it 
Leading order diagrams contributing to the deuteron anapole
  form factor.  These are the magnetic coupling contributions, written 
  as $A_M(|{\bf k}|)$  in the text.    
  The crossed circles represent operators that create  
  or annihilate two nucleons with
  the quantum numbers of the deuteron.
  The solid square is the weak operator with 
  coefficient $h^{(1)}_{\pi NN}$, and the open circle is the magnetic 
  coupling
  to the electromagnetic field.
  Wavy lines are photons, solid lines are nucleons, 
  and dashed lines are mesons.
  }
\label{fig:anam}
\vskip .2in
\end{figure}
\begin{figure}[t]
\centerline{\epsfxsize=4.0in \epsfbox{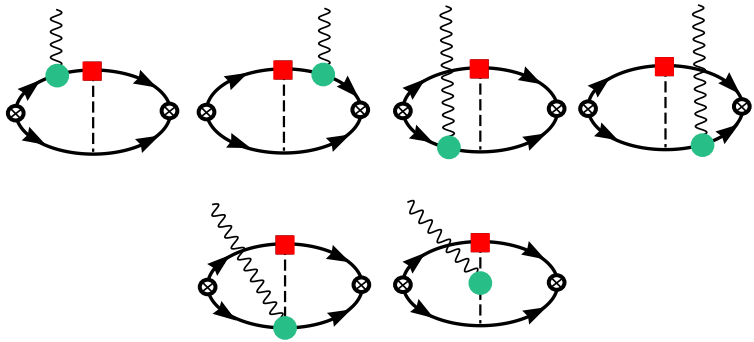}}
\noindent
\caption{\it Leading order diagrams contributing to the deuteron anapole
  form factor.  These are electric coupling contributions, written 
  as $A_E(|{\bf k}|)$  in the text.   
  The crossed circles represent operators that create  
  or annihilate two nucleons with
  the quantum numbers of the deuteron.
  The solid square is the weak operator with 
  coefficient $h^{(1)}_{\pi NN}$, and the solid circle is the minimal 
  coupling
  to the electromagnetic field.
  Wavy lines are photons, solid lines are nucleons, 
  and dashed lines are mesons.
  }
\label{fig:anael}
\vskip .2in
\end{figure}

The anapole form factor, $A_D(|{\bf k}|)$, is  separated into
three parts:  the single nucleon contribution $A_{N}$, the magnetic
contribution $A_{M}$, and the electric contribution $A_{E}$, with
\begin{eqnarray}
A_D(|{\bf k}|) = A_{N}(|{\bf k}|) + A_M(|{\bf k}|) + A_E(|{\bf k}|) 
\ \ .
\end{eqnarray}
The expressions we obtain for the form factors use
methods outlined in ref. \cite{binger}.  
While the regularization and renormalization prescription
used in \cite{binger} does not directly 
lead to gauge invariant results, 
by matching orders in a $|{\bf k}|^2$ expansion to known 
dimensionally regularized results, we can
recover the full dimensionally regulated expression.  
In the appendix we show individual integrals which are relevant for this
calculation (as well as other two loop pion exchange calculations).

The diagrams in fig.~(\ref{fig:anasn})
that constitute the contribution from the single nucleon
factorize into the product of the 
single nucleon anapole form factor 
(generated by $h_{\pi NN}^{(1)}$, see \cite{Musolfa}) 
with the LO deuteron charge form factor to yield
\begin{eqnarray}
A_{N}(|{\bf k}|) = - {
g_A e h_{\pi NN}^{(1)} M_N^2
\over \pi f} {2 \gamma \over |{\bf k}|^4} \left[
(m_\pi^2 + {|{\bf k}|^2 \over 4}) \  \tan^{-1} 
 \left ({|{\bf k}| \over 2 m_\pi} \right)  - { m_\pi |{\bf k}| \over 2} \right] 
\tan^{-1} \left( {|{\bf k}| \over 4 \gamma } \right) \ \ ,
\end{eqnarray}
where $\gamma = \sqrt{M_N B}$ is the binding momentum of the 
deuteron with $B$ the deuteron binding energy.

The magnetic contribution, arising from the interaction of the 
electromagnetic field with the isovector magnetic moment of the nucleon,
comes from the diagrams shown in 
fig.~(\ref{fig:anam}).
\begin{eqnarray}
A_M (|{\bf k}|) &=& -{
g_A e h_{\pi NN}^{(1)} M_N^2
\over 2 \pi  f} {\kappa_1 \over |{\bf k}|^2} 
\left[2 \gamma - m_\pi + {m^2_\pi \over 2 \gamma} \logone 
\right. \\ \nonumber
&&
+ \ {|{\bf k}| \over 4 \gamma^2} 
\left(m^2_\pi \log \left[1+{2 \gamma \over m_\pi }\right]
+ 2 \gamma^2-
2 \gamma m_\pi \right) \atanfour\  
\\ \nonumber
&& \ 
+ \ \left( -{8 \gamma^2 \over |{\bf k}|} - {|{\bf k}| \over 2}
+{ |{\bf k}| m_\pi \over 2 \gamma} + {2 m^2_\pi \over |{\bf k}|} \right) 
\atanthree 
\\ \nonumber
&& 
+\ \ i {m^2_\pi |{\bf k}| \over 8 \gamma^2} \left. 
\left(  Li_2 \left( {-i |{\bf k}| \over 2(m_\pi+2 \gamma)}\right) -
 Li_2 \left({i |{\bf k}| \over 2(m_\pi +2 \gamma)}\right) 
 \right.\right.
\\ \nonumber
&& \left.\left. \qquad\qquad \qquad
-  Li_2 \left({-i |{\bf k}| -4 \gamma \over 2 m_\pi}\right) + 
 Li_2 \left({i |{\bf k}| -4 \gamma \over 2 m_\pi}\right)
 \   \right)  \right] \ \ , 
\end{eqnarray}
where $Li_2(z)$ is the dilogarithm function, given in the appendix.

The electric contribution, arising from the minimal coupling of the 
electromagnetic field to pions and non-relativistic nucleons,  
comes from the diagrams shown in
fig.~(\ref{fig:anael}). 
\begin{eqnarray}
A_E (|{\bf k}|) &=& -{g_A e h_{\pi NN}^{(1)} M_N^2 \over 8 \pi f}\
 {1\over |{\bf k}|  }  
\left\{
-m_\pi |{\bf k}|
- {(m_\pi-\gamma)( |{\bf k}|^2+16\gamma^2)\over 2\gamma}  
\tan^{-1} \left({|{\bf k}| \over 4 \gamma}\right) 
\right.
\nonumber\\
& & \left.
+ { |{\bf k}|^2 (m_\pi-\gamma) + 4\gamma (m_\pi^2-4\gamma^2)\over 2\gamma}
 \tan^{-1}\left({|{\bf k}| \over 2(m_\pi +2 \gamma)}\right)
\right.
\nonumber\\
& & \left.
- { |{\bf k}| m_\pi^2\over 2\gamma} 
\log\left[{{|{\bf k}|^2 \over 4} + (m_\pi +2\gamma)^2\over (m_\pi +2\gamma)^2}\right]
\ +\ { m_\pi^2 ( |{\bf k}|^2+16\gamma^2)\over 4\gamma^2}
\tan^{-1} \left({|{\bf k}| \over 4 \gamma}\right) 
\log\left[1+{2\gamma\over m_\pi}\right]
\right.
\nonumber\\
& & \left.
- 2\gamma\rho  \tan^{-1} \left({|{\bf k}| \over 2m_\pi +\rho}\right)
\log\left[{|{\bf k}|^2 + 4 (m_\pi +2\gamma)^2\over (\rho +4\gamma)^2}\right]
\right.
\nonumber\\
& & \left.
+ 2\gamma\rho  \tan^{-1} \left({|{\bf k}| \over 2(m_\pi+2\gamma)}\right)
\log\left[ { (\rho-4\gamma)^2\over2\rho (\rho+2m_\pi)}\right]
\right.
\nonumber\\
& &\left.
+ i {m_\pi^2  ( |{\bf k}|^2+16\gamma^2) \over 8\gamma^2}
\left[
       Li_2\left({i |{\bf k}| -4\gamma \over 2 m_\pi } \right)
- 
 Li_2\left({-i |{\bf k}| -4\gamma \over 2 m_\pi } \right)
\right.
\right.
\nonumber\\
& &
\left.\left.
\qquad\qquad\qquad\qquad
    +   Li_2 \left({-i |{\bf k}| \over 2(m_\pi +2\gamma)}\right)
- Li_2 \left({i |{\bf k}| \over 2(m_\pi +2\gamma)}\right)
\right]
\right.
\nonumber\\
& & \left.
+ i 2\gamma\rho 
\left[ 
Li_2 \left({-i |{\bf k}| - 2 m_\pi - 4\gamma \over \rho-4\gamma}\right)
- 
Li_2 \left({i |{\bf k}| - 2 m_\pi - 4\gamma \over \rho-4\gamma}\right)
\right.\right.
\nonumber\\
& & 
\left.\left.
\qquad\qquad
+
Li_2 \left({-i |{\bf k}| + \rho - 2 m_\pi \over \rho+4\gamma}\right)
-
Li_2 \left({i |{\bf k}| + \rho - 2 m_\pi \over \rho+4\gamma}\right)
\right]
\ \ 
\right\}
\ \ ,
\end{eqnarray}
where $\rho= \sqrt{|{\bf k}|^2 + 4 m^2_\pi}$.

%
\begin{figure}[t]
\centerline{{\epsfxsize=4.5in \epsfbox{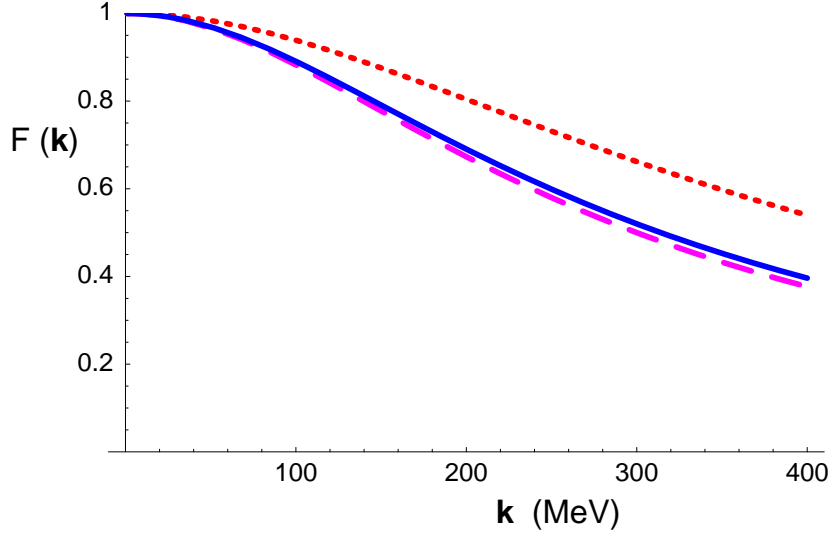}} }
\noindent
\caption{\it
The three contributions to the anapole form factor  as a
function of momentum transfer $|{\bf k}|$, each normalized to unity at 
$|{\bf k}|=0$. The solid curve corresponds to the contribution
from the single nucleon anapole form factor, 
$\tilde{A}_N(|{\bf k}|)$ (see fig.~(\ref{fig:anasn})).
The dashed curve shows the contribution
from the photon coupling to the isovector nucleon magnetic moment, 
$\tilde{A}_M(|{\bf k}|)$ (see fig.~(\ref{fig:anam})).
The dotted curve corresponds to the electric form factor, 
$\tilde{A}_E(|{\bf k}|)$
(see fig.~(\ref{fig:anael})).
}
\label{fig:FFplot}
\vskip .2in
\end{figure}

\section{Discussion of Results}

For illustrative purposes, we explicitly factor out the normalization
of each of the form factors contributing to the anapole moment of the 
deuteron. 
The normalized 
single nucleon, magnetic, and electric form factors,
$\tilde A_N$, $\tilde A_M$, and $\tilde A_E$ respectively, are
defined through
\begin{eqnarray}
        A_D(|{\bf k}|) & = & -{e g_A h_{\pi NN}^{(1)} M_N^2 \over 12 \pi f}
\left[ \ {1\over 2 m_\pi} 
\tilde A_N (|{\bf k}|) \ + \ 
\kappa_1\ { m_\pi+\gamma\over (m_\pi+2\gamma)^2}\ 
\tilde A_M (|{\bf k}|)
\right.\nonumber\\
& & \left. \qquad\qquad\qquad\qquad
\ -\  { m_\pi^2+3 m_\pi\gamma + 12\gamma^2\over 6 m_\pi (m_\pi+2\gamma)^2}
\tilde A_E (|{\bf k}|)\ \ 
\right] 
\nonumber\\
& = & 
-{e g_A h_{\pi NN}^{(1)} M_N^2 \over 12 \pi f}
\left[ \ {1\over 2 m_\pi} \ + \
\kappa_1\ { m_\pi+\gamma\over (m_\pi+2\gamma)^2}\ 
-{ m_\pi^2+3 m_\pi\gamma + 12\gamma^2\over 6 m_\pi (m_\pi+2\gamma)^2}
\right] 
\tilde A_D (|{\bf k}|)
\ \ \ ,
\label{eq:anaaX}
\end{eqnarray}
where $\tilde A_M (0)\ =\ \tilde A_N (0)\ =\ \tilde A_E (0)\ = \ 1$.
The deuteron anapole form factor,  $\tilde A_D  (|{\bf k}|)$,
is normalized so that  $\tilde A_D  (0)=1$.  (Note that while
$A_D=A_N+A_M+A_E$, the normalized $\tilde A_D$ does {\sl not} equal
the sum of the individual tilded form factors.)

Fig.~(\ref{fig:FFplot}) shows that at LO
in the effective field
theory expansion the form factor associated with the single nucleon
and magnetic contributions fall somewhat faster than the contribution from the 
electric form factor, versus increasing
momentum transfer.  The power counting for this effective field theory
\cite{KSW} suggests that corrections 
to the LO result are naively
at the 30\% level.
While it is possible that the differences between the form factors computed
at LO are modified by higher order contributions, there is no reason for the 
form factors to be identical.

The anapole form factor of the deuteron contributes to parity violation
in electron scattering off the deuteron.  A simultaneous source of
parity violation in such an experiment is provided by any intrinsic
strangeness existing in the deuteron.  In a previous calculation 
\cite{SSana}, we
gave an expression for the leading order parity violating matrix element in
electron-deuteron scattering in terms of the deuteron anapole
moment and the strange axial matrix element of the nucleon.  
The
present calculation extends this work, and increases the possibility
of separating the two effects, by generalizing the moment to 
the momentum dependent form factor.  To do this we need to estimate
how the strange axial matrix element depends on momentum transfer.

The scale for the variation of the anapole form
factors is set by the deuteron binding momentum, $\gamma$, and the pion mass.
In contrast, the variation of the form factors describing the matrix element 
of the strange axial current is set by the deuteron binding momentum, and 
the kaon mass or some higher mass scale.
Therefore, we expect the contribution to parity violating electron-deuteron
scattering from the matrix element of the axial strange quark current to 
become more important at higher momentum transfers.
The fall of the strange quark form factor with momentum should be driven by
the small deuteron binding energy; that is, by the nucleon recoil effects.
This gives 
the LO strange deuteron form factor
the same structure as the LO deuteron charge form factor,
\begin{eqnarray}
F_C(|{\bf k}|) & = & {4\gamma\over |{\bf k}|}
\tan^{-1}\left( {|{\bf k}|\over  4\gamma}\right)
\ \ \ .
\label{eq:charge}
\end{eqnarray}
The parity violating amplitude in electron-deuteron scattering
is then, at LO in the effective field theory,
\begin{eqnarray}
  {\cal A}^{(pv)}(|{\bf k}|) & = &
  10^{-7} \left( 5.4\   g_{\Delta S} \ F_C (|{\bf k}|)
\ -\ 0.21 \left( {h^{(1)}_{\pi NN}\over  10^{-7}}\right) 
\tilde A_D (|{\bf k}|)
\right)
 {1\over M_N^2} \ \epsilon^{abi}\ 
 \overline{e}\gamma_i e\ 
 \varepsilon^*_a\ \varepsilon_b
 \ \ \ .
 \label{eq:pvmatX}
\end{eqnarray}

\begin{figure}[t]
\centerline{{\epsfxsize=4.5in \epsfbox{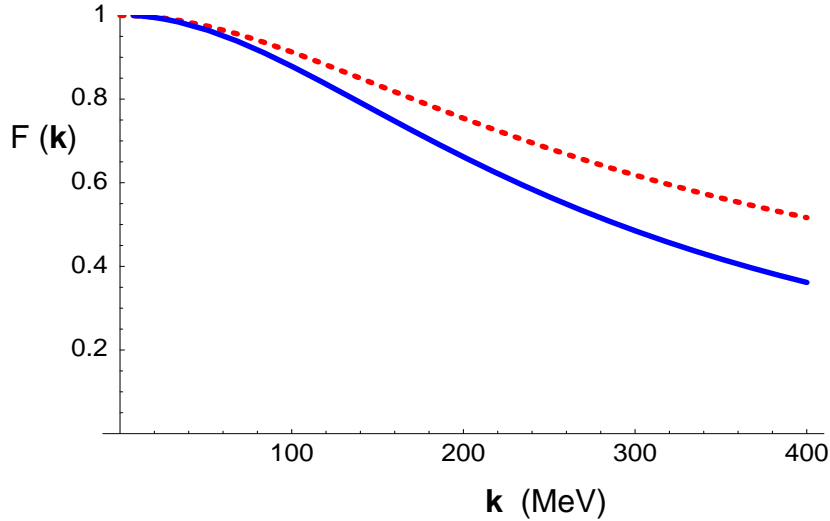}} }
\noindent
\caption{\it
A comparison between the normalized
anapole form factor,  $\tilde A_D(|{\bf k}|)$ defined in 
eq.~(\ref{eq:anaaX}), and the 
charge form factor, expected to describe the matrix element of the 
strange axial current at LO.
The solid curve is the anapole form factor while the dotted 
curve corresponds to the charge form factor $F_C(|{\bf k}|) $. 
}
\label{fig:FFcomp}
\vskip .2in
\end{figure}

The two contributions to the anapole form factor are shown in 
fig.~(\ref{fig:FFcomp}). 
%
%
A more precise estimate of the form factors, 
and hence a more precise estimate of $h_{\pi NN}^{(1)}$ through
a comparison with electron-deuteron scattering data, 
would require a higher order calculation of the anapole 
form factor in addition to more insight into possible strange quark
effects (while current measurements of $g_{\Delta s}$ are consistent with
zero\cite{SMC,EE,WSa}, so is one of the measurements of 
$h_{\pi NN}^{(1)}$\cite{Fluorine}.)
An estimate 
of the size of the higher order corrections entering at NLO and higher
can be made by examining the size of the corrections to the charge 
form factor,  as presented in \cite{KSW2}.
Compact, analytic expressions for the LO form factors have been obtained in
this work, and it is likely that similarly
simple expressions can be obtained at NLO.

It is appropriate to comment on the recent work by 
Khriplovich and Korkin\cite{KK} who determine the anapole
moment of the deuteron in the zero-range potential limit.
They find (using their revised expression for the  magnetic
moment term)
a contribution from the electric and magnetic interactions (not
including the single nucleon contribution) proportional to 
\begin{eqnarray}
A_D^{(KK)} & \propto & {m_\pi+\gamma\over (m_\pi+2\gamma)^2}
\left(\kappa_1 - {1\over 6}\right)
\ \ \ .
\label{eq:KK}
\end{eqnarray}
The expression we obtained in \cite{SSana} does not reproduce their result,
except in the limiting case where $\gamma\rightarrow 0$.
It is clear that their expression, shown in eq.~(\ref{eq:KK}), 
does not have the correct behavior in the chiral limit for  
fixed $\gamma$ (the deuteron anapole moment would diverge, driven by the
single nucleon contribution), as discussed in \cite{SSana}.  
The deuteron anapole moment cannot have any $1/m_\pi$ dependence
in the chiral limit when $\gamma$ is held fixed, due to the vanishing of the 
strong one-pion-deuteron coupling.
Our result, shown in eq.~(\ref{eq:anaaX}), correctly reproduces this 
necessary behavior.

\section{Conclusion}

In this work we have presented an analytic expression for the anapole
form factor at LO in the effective field theory expansion.
The dominant contribution to this form factor is from  pion physics, 
through the weak pion-nucleon coupling $h_{\pi NN}^{(1)}$.
The pion mass and
deuteron binding momentum are the scales that 
drive its variation.
In contrast, pion effects will not dominate the matrix element of the strange
axial current.
We find that while the deuteron binding momentum provides most of the 
momentum
variation of both form factors, the anapole form factor falls fasters than 
the matrix element of the strange axial current.
This has important implications for the determination of both $g_{\Delta s}$
and $h_{\pi NN}^{(1)}$ from parity violating electron-deuteron scattering.
Encouraged by the success of effective field theory for this parity
violating process, and the relatively simple closed form expressions we
obtain, we are optimistic about higher order calculations of this
form factor and the possibilities of extracting $h_{\pi NN}^{(1)}$ from
electron scattering experiments.

\vskip .5cm

This work is supported in part by the U.S. Dept. of Energy under
Grants No. DE-FG03-97ER4014 and DE-FG02-96ER40945, and NSF grant
number 9870475.

\vfill\eject


\section{Appendix of Integrals}
The analytic expressions we found for the anapole form factors
come from integrals which are relevant for
other calculations.
In this appendix we give expressions for the general
form of some of the 
integrals encountered in evaluating the Feynman diagrams for
the anapole calculation.
We use a combination of dimensional regularization and 
the position space representation (as advocated by 
Binger\cite{binger}) to find expressions for the integrals.
The divergent contribution
and well defined finite piece will be given in terms of
\begin{eqnarray}
\Gamma & = &  \Gamma[4-n]\  + \log\left(\pi\right)\ +\ 1 
\ \ \ ,
\label{eq:divdef}
\end{eqnarray}
where $n$ is the number of space-time dimensions. 
To be consistent with the PDS procedure,
$\mu/2$ and not $\mu$ is used for the dimensional regularization
mass-scale in the definition of the integrals.
In the expressions that follow,
the magnitude of the three-momentum transfer
is given by  $k=|{\bf k}|$. All integrals have been power-series
expanded about 
$n=4$, so terms of order $(n-4)^1$ or higher are not shown.
Where needed to distinguish one integral from another within
a class (same denominators in the integrand), 
a subscript is used to indicate the naive degree of ultraviolet
divergence of the diagram.


\subsection{1-Loop, 2-Propagators}

We find that two-loop integrals often reduce down to some
simple functions, dilogarithms, and the following integral
(which also appears in one loop diagrams):
\begin{eqnarray}
A(a,b;k) & = & \int {d^3 {\bf q}\over (2\pi)^3} 
{1\over [ ({\bf q}-{1\over 2}{\bf k})^2+a^2]}
{1\over [ {\bf q}^2+b^2]}
\ =\  
{1\over 2\pi k} \tan^{-1}\left({k\over 2(a+b)}\right)
\ \ \ .
\label{eq:tdef}
\end{eqnarray}

\subsection{2-Loops, 3-Propagators}


A frequently encountered integral in the theory with 
perturbative pions\cite{KSWa,KSW} is $B_0(a,b,c;\mu;k)$
evaluated at $k=0$ (an integral that is 
ultra-violet divergent as $n\rightarrow 4$)
where
\begin{eqnarray}
B_0(a,b,&&c;\mu; k)  
\nonumber \\
&& =
\left({\mu\over 2}\right)^{8-2n}
\int {d^{n-1} {\bf q}\over (2\pi)^{n-1}} 
{d^{n-1} {\bf l}\over (2\pi)^{n-1}} \ 
{1\over [ ({\bf q}-{1\over 2}{\bf k})^2+a^2]}
{1\over [ ({\bf q}+{\bf l})^2+b^2]}
{1\over [ {\bf l}^2+c^2]}
\ \ \ .
\label{eq:yk0}
\end{eqnarray}
It can be computed at $k=0$ using dimensional regularization:
\begin{eqnarray}
B_0(a,b,c;\mu; 0) & = & -{1\over (4\pi)^{n-1}}\ 
\Gamma [4-n]\  \Gamma [{3-n\over 2}]\  \Gamma [{n-3\over 2}]\ 
\left( {2 M\over \mu}\right)^{ 2n-8}
\nonumber\\
& = & 
{1\over 32\pi^2} \left[ \Gamma
-2 \log\left[{M\over\mu}\right]
\ \ \right]
\ \ \ ,
\label{eq:y00}
\end{eqnarray}
where $M=a+b+c$.
The methods of \cite{binger} are used to find the finite $k$-dependent portion, which 
matches onto the previous result to give the full expression 
for arbitrary $k$:
\begin{eqnarray}
B_0(a,b,c;\mu;k) & = & 
{1\over 32\pi^2}
\left[\  
\Gamma -
\log\left[{M^2+{k^2\over 4}\over\mu^2}\right]
- {4 M\over k}\tan^{-1}\left({k\over 2 M}\right)
+2 
\  \right]
\ \ \ .
\label{eq:yk}
\end{eqnarray}
This reduces to eq.~(\ref{eq:y00}) in the limit $k\rightarrow 0$, as required.


\noindent A related integral is
\begin{eqnarray}
B_1(a,b,c;\mu;k) & = & 
\left({\mu\over 2}\right)^{8-2n}
\int {d^{n-1} {\bf q}\over (2\pi)^{n-1}} 
{d^{n-1} {\bf l}\over (2\pi)^{n-1}} \ 
{ {\bf q}\cdot {\bf k}\over [ ({\bf q}-{1\over 2}{\bf k})^2+a^2]}
{1\over [ ({\bf q}+{\bf l})^2+b^2]}
{1\over [ {\bf l}^2+c^2]}
\ .
\label{eq:xk0}
\end{eqnarray}
Making the replacement 
${\bf q}\cdot {\bf k} \rightarrow -({\bf q}-{1\over 2}{\bf k})^2 
+ {\bf q}^2 + {1\over 4}{\bf k}^2$ 
leaves a more divergent integral (the ${\bf q}^2$ term) 
than found in the $B_0$ integral,
but a judicious combination of terms allows 
explicit evaluation of this integral using
the methods of \cite{binger}.
We find
\begin{eqnarray}
B_1(a,b,c;\mu;k) =  
{1\over 32\pi^2} & & 
\left[ \ 
 {k^2\over 3}
\left( \Gamma + 2 - \log\left[{M^2+{k^2\over 4}\over\mu^2}\right]\right)
\ +\  2 a M 
-\ {2\over 3}M^2
\right.\nonumber\\
 & & \left.
-\ 
\left( {4a\over k} M^2+M k - {4\over 3k} M^3 + a k\right) 
\tan^{-1}\left({k\over 2 M}\right)
\right]
\ \ \ .
\label{eq:xk}
\end{eqnarray}

\subsection{2-Loops, 4-Propagators}

In graphs where an external current couples to a nucleon line
and a perturbative pion is exchanged between nucleons, we 
encounter integrals with four propagators and  two loop momentum integrations.
Such integrals are expressed in terms of simple functions and dilogarithmic
functions.
The function $P(X,Y;k)$ is defined to be
\begin{eqnarray}
P(X,Y;k) & = & \int_{-1}^{+1}\ dz\ 
{\log\left[ X + i {k z\over 2}\right]\over Y + i {k z\over 2}}
\nonumber\\
& = &
 -i {2\over k} 
\left\{ 
Li_2\left[{X+ i {k\over 2} + i\epsilon\over X-Y+ i\epsilon}\right]
\ -\ 
Li_2\left[{X- i {k\over 2} - i\epsilon\over X-Y- i\epsilon}\right]
\right. \nonumber \\
&& + i 2 \pi \log\left[X-Y+i \epsilon\right] 
\ \theta\left[X-Y\right]\theta\left[Y\right]
\nonumber\\
& &
+ \log\left[X+ i {k\over 2} + i\epsilon \right] 
\log\left[{Y+ i {k\over 2} + i\epsilon\over Y-X + i\epsilon}\right]
\nonumber\\
&& \left.
- \log\left[X- i {k\over 2} - i\epsilon \right] 
\log\left[{Y- i {k\over 2} - i\epsilon\over Y-X - i\epsilon}\right]
\ \right\}
\ \ \ ,
\label{eq:Pdef}
\end{eqnarray}
where the dilogarithmic function $Li_2(z)$ is conventionally defined
\begin{eqnarray}
Li_2 (z) & = & -\int_0^z\ dt \ {\log[1-t]\over t}
\ =\  \int_0^1\ dx\  { \log[x]\over x-{1\over z}}
\ \ \ .
\label{eq:dilog}
\end{eqnarray}


The simplest integral involving four propagators at two loops
does not have any loop momentum dependence in the numerator, and is
finite at $n=4$.
It is expressed in terms of functions defined previously,
\begin{eqnarray}
& & I_{-2} (a,b,c,d;k) \ =\ 
\int {d^{3} {\bf q}\over (2\pi)^{3}} 
{d^{3} {\bf l}\over (2\pi)^{3}} \ 
{1\over [ ({\bf q}-{1\over 2}{\bf k})^2+a^2]}
{1\over [ {\bf q}^2+b^2]}
{1\over [ ({\bf q}+{\bf l})^2+c^2]}
{1\over [ {\bf l}^2+d^2]}
\nonumber\\
 & & =  
{1\over 64\pi^2 b}
\left\{ \phantom{{k \over k}} P(a+c+d,a-b;k) - P(a+c+d,a+b;k) 
\right. \nonumber\\
& & \left.\quad
+ i {2\over k} \log\left(b+c+d\right)
\left[ 
\log\left(a-b+i {k\over 2} + i \epsilon\right)
- 
\log\left(a-b-i {k\over 2} - i \epsilon\right)
- i 2 \pi \theta(b-a)
\right.\right.\nonumber\\
& & \left.\left.\qquad\qquad\qquad\qquad\qquad
-
\log\left(a+b+i {k\over 2} + i \epsilon\right)
+
\log\left(a+b-i {k\over 2} - i \epsilon\right)
\right]\ 
\right\}
\ \ \ .
\label{eq:I0answ}
\end{eqnarray}


An integral with a more complicated momentum dependence, but  without Lorentz structure in
the numerator of the integrand, can be written in terms of $I_{-2}$,
\begin{eqnarray}
\tilde I_{-2}(a,b,c,d;k) 
& =&  
\int {d^{3} {\bf q}\over (2\pi)^{3}} 
{d^{3} {\bf l}\over (2\pi)^{3}} \ 
{1\over [ ({\bf q}-{1\over 2}{\bf k})^2+a^2]}
{1\over [ ({\bf q}+{1\over 2}{\bf k})^2+b^2]}
{1\over [ ({\bf q}+{\bf l})^2+c^2]}
{1\over [ {\bf l}^2+d^2]}
\nonumber\\
& = & 
{1\over 2}
\left[ I_{-2}(a, \sqrt{ {k^2\over 4} + {a^2\over 2} + {b^2\over 2}} ,c,d)
\ +\ 
 I_{-2}(b, \sqrt{ {k^2\over 4} + {a^2\over 2} + {b^2\over 2}} ,c,d)
\right]
\ \ \ .
\end{eqnarray}


We will use the following integrals in subsequent expressions.  
\begin{eqnarray}
I_{-1}(a,b,c,d;k) & = &
 \int {d^3 {\bf q}\over (2\pi)^3} {d^3 {\bf l}\over (2\pi)^3} 
{{\bf q}\cdot {\bf k}\over [ ({\bf q}-{1\over 2}{\bf k})^2+a^2]}
{1\over [ {\bf q}^2+b^2]}
{1\over [ ({\bf q}+{\bf l})^2+c^2]}
{1\over [ {\bf l}^2+d^2]}
\nonumber\\
& = & B_0(a,c,d;\mu;k) - B_0(b,c,d;\mu;0) - (b^2-a^2-{k^2\over 4}) I_{-2}(a,b,c,d;k)
\ \ \ ,
\end{eqnarray}
\begin{eqnarray}
I_0(a,b,c&&,d;\mu;k) 
\nonumber\\
&& = \left({\mu\over 2}\right)^{8-2n}
\int {d^{n-1} {\bf q}\over (2\pi)^{n-1}} 
{d^{n-1} {\bf l}\over (2\pi)^{n-1}} \ 
{\left({\bf q}\cdot {\bf k}\right)^2 
\over [ ({\bf q}-{1\over 2}{\bf k})^2+a^2]}
{1\over [ {\bf q}^2+b^2]}
{1\over [ ({\bf q}+{\bf l})^2+c^2]}
{1\over [ {\bf l}^2+d^2]}
\nonumber\\
&&=
B_1(a,c,d;\mu;k) + 
(b^2-a^2-{k^2\over 4})\  \left( B_0(b,c,d;\mu;0)-B_0(a,c,d;\mu;k)\right)
\nonumber\\
&& \ \ 
+ 
(b^2-a^2-{k^2\over 4})^2\  I_{-2}(a,b,c,d;k)
\ \ \ .
\end{eqnarray}


The first integral with a Lorentz index, but
finite at $n=4$, is
\begin{eqnarray}
J_{-1}^j (a,b,&&c,d;k)  =  
\int {d^3 {\bf q}\over (2\pi)^3} {d^3 {\bf l}\over (2\pi)^3} 
{{\bf l}^j\over [ ({\bf q}-{1\over 2}{\bf k})^2+a^2]}
{1\over [ {\bf q}^2+b^2]}
{1\over [ ({\bf q}+{\bf l})^2+c^2]}
{1\over [ {\bf l}^2+d^2]}
\nonumber\\
&& =  {{\bf k}^j\over {\bf k}^2}
\int {d^3 {\bf q}\over (2\pi)^3} {d^3 {\bf l}\over (2\pi)^3} 
{1\over [ {\bf q}^2 ]}
{({\bf q}\cdot {\bf k})({\bf q}\cdot {\bf l})\over 
[ ({\bf q}-{1\over 2}{\bf k})^2+a^2]}
{1\over [ {\bf q}^2+b^2]}
{1\over [ ({\bf q}+{\bf l})^2+c^2]}
{1\over [ {\bf l}^2+d^2]}
\ \ \ ,
\label{eq:J0def}
\end{eqnarray}
where we have used the known external momentum dependence to 
rewrite the ${\bf l}$ integrand 
in terms of the ${\bf q}$ variable.  
After completing squares in the numerator we obtain 
\begin{eqnarray}
J_{-1}^j &&(a,b,c,d;k)
\nonumber \\
&&= \ - { {\bf k}^j\over 2\ k^2}
\left\{
\left({c-d\over 4\pi}\right)
\left[ A(0,b;0)-A(a,b;k) \
- \left({a^2+{1\over 4}k^2\over b^2}\right)\left(A(a,0;k)-A(a,b;k)\right)
\right]
\right.\nonumber\\
&&+ B_0(a,c,d;\mu;k)-B_0(b,c,d;\mu;0)
- \left(b^2-a^2-{1\over 4}k^2\right) I_{-2}(a,b,c,d;k)
\nonumber\\
&&+\left(d^2-c^2\right)
\left[ \phantom{{k \over k}} I_{-2}(0,b,c,d;0)-I_{-2}(a,b,c,d;k)
\right.\nonumber\\
& & \left.\left.
\qquad\qquad
- \left({a^2+{1\over 4}k^2\over b^2}\right)
\left(I_{-2}(a,0,c,d;k)-I_{-2}(a,b,c,d;k)\right)\right] \
\right\}
\ \ \ .
\label{eq:J0}
\end{eqnarray}


The following integrals have two indices in the numerator of the integrand.
These again are naively divergent and must be treated in $n$ dimensional
space-time.
The divergences all enter through integrals already defined, $B_0$, $B_1$,
$I_{-2}$, and $I_0$.

\begin{eqnarray}
K^{i,j} & = &
\left({\mu\over 2}\right)^{8-2n}
\int {d^{n-1} {\bf q}\over (2\pi)^{n-1}} 
{d^{n-1} {\bf l}\over (2\pi)^{n-1}} \ 
{{\bf q}^i\ {\bf l}^j\over 
[ ({\bf q}-{1\over 2}{\bf k})^2+a^2]}
{1\over [ {\bf q}^2+b^2]}
{1\over [ ({\bf q}+{\bf l})^2+c^2]}
{1\over [ {\bf l}^2+d^2]}
\nonumber\\
& =  &\ 
 {1\over n-2}\left( K_0 - \tilde{K}_0 \right){\bf \delta}^{ij}
\ +\ 
{1\over n-2} \left( (n-1) \tilde{K}_0- K_0\right)  
{{\bf k}^i {\bf k}^j\over {\bf k}^2}
\ \ \ ,
\label{eq:L0n}
\end{eqnarray}
where
\begin{eqnarray}
&&K_0 (a,b,c,d;\mu;k) 
\nonumber \\
&& =
\left({\mu\over 2}\right)^{8-2n}
\int {d^{n-1} {\bf q}\over (2\pi)^{n-1}} 
{d^{n-1} {\bf l}\over (2\pi)^{n-1}} \ 
{{\bf q}\cdot {\bf l}\over 
[ ({\bf q}-{1\over 2}{\bf k})^2+a^2]}
{1\over [ {\bf q}^2+b^2]}
{1\over [ ({\bf q}+{\bf l})^2+c^2]}
{1\over [ {\bf l}^2+d^2]} 
\nonumber\\
& & = {1\over 2}
\left[\ \ 
\left({c-d\over 4\pi}\right) A(a,b;k)
\ -\ B_0(a,c,d;\mu;k)
\ +\ \left(d^2+b^2-c^2\right) I_{-2}(a,b,c,d;k)\ \ 
\right]
\ \ \ ,
\label{eq:L01n}
\end{eqnarray}
and 
\begin{eqnarray}
&&\tilde{K}_0 (a,b,c,d;\mu;k) 
\nonumber \\
&&=
{1\over {\bf k}^2}
\left({\mu\over 2}\right)^{8-2n}
\int {d^{n-1} {\bf q}\over (2\pi)^{n-1}} 
{d^{n-1} {\bf l}\over (2\pi)^{n-1}} \ 
{({\bf q}\cdot {\bf k})({\bf l}\cdot {\bf k})\over 
[ ({\bf q}-{1\over 2}{\bf k})^2+a^2]}
{1\over [ {\bf q}^2+b^2]}
{1\over [ ({\bf q}+{\bf l})^2+c^2]}
{1\over [ {\bf l}^2+d^2]}
\nonumber\\
& & =
{1\over 2 k^2}
\left[
\left({d-c\over 4\pi}\right)
\left( {a-b\over 4\pi} + (b^2-a^2-{k^2\over 4}) A(a,b;k)\right)
- B_1(a,c,d;\mu;k) 
\right.\nonumber\\
& & - \left. 
(d^2+b^2-c^2)\left( B_0(b,c,d;\mu;0) - B_0(a,c,d;\mu;k) + (b^2-a^2-{k^2\over 4})
  I_{-2}(a,b,c,d;k)\right)
\right.\nonumber\\
& & + \left. 
\left(a^2+{k^2\over 4}\right)
\left[ 
\left({d-c\over 4\pi}\right)\left( A(0,b;0)-A(a,b;k) 
- {a^2+{k^2\over 4}\over b^2} \left(A(a,0;k)-A(a,b;k)\right)\right)
\right.\right.\nonumber\\
& & \left.\left. 
\qquad\qquad +
B_0(0,c,d;\mu;0) - B_0(a,c,d;\mu;k) - ({a^2+{k^2\over 4}}) I_{-2}(a,0,c,d;k)
\right.\right.\nonumber\\
& &  \left.\left. 
\qquad\qquad -
(d^2+b^2-c^2)\left[ 
{1\over b^2}\left( B_0(0,c,d;\mu;0)-B_0(b,c,d;\mu;0)\right) 
-I_{-2}(a,b,c,d;k) 
\right.\right.\right.\nonumber\\
& &  \left.\left.\left. 
\qquad\qquad\qquad\qquad -
{a^2+{k^2\over 4}\over b^2}\left( I_{-2}(a,0,c,d;k)-I_{-2}(a,b,c,d;k)\right)
\right]\ 
\right]\ \ 
\right]
\ \ \ .
\label{eq:L02n}
\end{eqnarray}
It is necessary to keep the $n$-dependence of the coefficients in the tensor
structure of eq.(\ref{eq:L0n}) explicit, since factors of $n-4$  combine with 
divergences in the $K_0$ and $\tilde{K}_0$ to produce finite contributions. 
These are necessary to recover a gauge invariant result for the 
weak one-photon matrix element in the deuteron, and hence the anapole moment.

Finally we present another two index integral
\begin{eqnarray}
L^{i,j}  & =&  
\left({\mu\over 2}\right)^{8-2n}
\int {d^{n-1} {\bf q}\over (2\pi)^{n-1}} 
{d^{n-1} {\bf l}\over (2\pi)^{n-1}}
\nonumber \\
&& \qquad \qquad \qquad
{{\bf q}^i\ {\bf q}^j\over 
[ ({\bf q}-{1\over 2}{\bf k})^2+a^2]}
{1\over [ ({\bf q}+{1\over 2}{\bf k})^2+b^2]}
{1\over [ ({\bf q}+{\bf l})^2+c^2]}
{1\over [ {\bf l}^2+d^2]}
\nonumber\\
 & =&  \ 
 {1\over n-2}\left( L_0 - \tilde L_0\right){\bf \delta}^{ij}
\ +\ 
{1\over n-2} \left( (n-1) \tilde L_0- L_0\right)  
{{\bf k}^i {\bf k}^j\over {\bf k}^2}
\ \ \ ,
\label{eq:L1n}
\end{eqnarray}
where
\begin{eqnarray}
& & L_0(a,b,c,d;\mu;k) 
\nonumber \\
&&=
\left({\mu\over 2}\right)^{8-2n}
\int {d^{n-1} {\bf q}\over (2\pi)^{n-1}} 
{d^{n-1} {\bf l}\over (2\pi)^{n-1}} \ 
{{\bf q}^2\over 
[ ({\bf q}-{1\over 2}{\bf k})^2+a^2]}
{1\over [ ({\bf q}+{1\over 2}{\bf k})^2+b^2]}
{1\over [ ({\bf q}+{\bf l})^2+c^2]}
{1\over [ {\bf l}^2+d^2]} 
\nonumber \\
&&=
{1\over 2}
\left[
B_0(b,c,d;\mu;k) + B_0(a,c,d;\mu;k) 
- (b^2+a^2+{k^2\over 2}) \tilde I_{-2}(a,b,c,d;k)
\right] 
\ \ \ ,
\label{eq:L11def}
\end{eqnarray}
and
\begin{eqnarray}
& & \tilde L_0(a,b,c,d;\mu;k)
\nonumber\\  
&  & =
{1\over {\bf k}^2}
\left({\mu\over 2}\right)^{8-2n}
\int {d^{n-1} {\bf q}\over (2\pi)^{n-1}} 
{d^{n-1} {\bf l}\over (2\pi)^{n-1}} \ 
{({\bf q}\cdot {\bf k})^2\over 
[ ({\bf q}-{1\over 2}{\bf k})^2+a^2]}
{1\over [ ({\bf q}+{1\over 2}{\bf k})^2+b^2]}
{1\over [ ({\bf q}+{\bf l})^2+c^2]}
{1\over [ {\bf l}^2+d^2]} 
\nonumber \\
&&=
{1\over 2 k^2}
\left[ I_{0}(a, \sqrt{ {k^2\over 4} + {a^2\over 2} + {b^2\over 2}},c,d;\mu;k)
+ 
 I_{0}(b,\sqrt{ {k^2\over 4} + {a^2\over 2} + {b^2\over 2}},c,d;\mu;k)
\right]
\ \ \ .
\label{eq:L12def}
\end{eqnarray}

We have shown the set of integrals needed to calculate
the anapole form factor of the deuteron.  We see that the diagrams
involved, even with finite momentum 
transfer  from external currents, are computible in closed form.
These results will be useful for many other two-loop computations.

\end{document}